\input harvmac

\noblackbox
\writedefs
\writetoc

\catcode`\@=11
\def\em@rk{\hbox{}} 
\def\xeqn{\expandafter\xe@n}\def\xe@n(#1){#1}
\def\xeqna#1{\expandafter\xe@na#1}\def\xe@na\hbox#1{\xe@nap #1}
\def\xe@nap$(#1)${\hbox{$#1$}}
\def\eqns#1{(\e@ns #1{\hbox{}})}
\def\e@ns#1{\ifx\und@fined#1\message{eqnlabel \string#1 is undefined.}%
\xdef#1{(?.?)}\fi \edef\next{#1}\ifx\next\em@rk\def\next{}%
\else\ifx\next#1\xeqn#1\else\def\n@xt{#1}\ifx\n@xt\next#1\else\xeqna#1\fi
\fi\let\next=\e@ns\fi\next}
\catcode`\@=12

\lref\rBBIPZ{E. Br\'ezin, C. Itzykson, G. Parisi and J.-B. Zuber,
{\it Comm. Math. Phys.} 59 (1978) 35\semi Bessis, C. Itzykson, and J.-B.
Zuber, {\it Adv. Appl. Math.} 1 (1980) 109.}
\lref\rSt{ M. Staudacher, {\it Phys.Lett }B305 (1993)332.}
\lref\rDSBKGM{M. Douglas and S. Shenker, {\it Nucl. Phys.} B335 (1990)
635\semi E. Br\'ezin and V. Kazakov, {\it Phys. Lett.} B236 (1990) 144\semi
D. Gross and A. Migdal, {\it Phys. Rev. Lett.} 64 (1990) 127;
{\it Nucl. Phys.} B340 (1990) 333.}%
\lref\rMehta{M.L. Mehta, {\it Random matrices, second edition}, ed. by Academic Press inc (1991).}
\lref\rDKK{J.M. Daul, V.A. Kazakov, I.K. Kostov, {\it Nuc.Phys}B409 (1993) 311, {\it Rational theories of 2d gravity from the two-matrix model}}
\lref\rDyson{F.J. Dyson, {\it Commun. Math. Phys.} 19(1970) 235; {\it Correlations between the eigenvalues of a random matrix}.}
\lref\rWigner{E.P. Wigner, {\it Proc. Cambridge Philos. Soc.} 47(1951) 790, reprinted in {\it C.E. Porter, Statistical theories of spectra: fluctuations }(Academic Press, New York, 1991).}
\lref\rCamarda{H.S. Camarda, {\it Phys.Rev.}A45 (1992) 579.}
\lref\rHWB{G. Hacken, R. Werbin, J. Rainwater, {\it Phys.Rev.}C17 (1978)43; W.M. Wilson, E.G. Bilpuch, G.E. Mitchel, {\it Nucl.Phys.}A245(1975); R.U. Hag, A. Pandey, O. Bohigas, {\it Phys.Rev.Lett.}48(1982) 1086.}
\lref\rCG{H.S. Camarda, P.D. Georgopulos, {\it Phys.Rev.Lett.}50(1983) 492.}
\lref\rBrZe{E. Br\'ezin and A. Zee, {\it Nucl. Phys.}B402 (1993); {\it Correlation functions in disordered systems}.}
\lref\rMS{M.L. Mehta, P. Shukla, {\it Two coupled matrices: Eigenvalue correlations and spacing functions}. {\it J. Phys.} A: Math Gen. 27 (1994) 7793-7803.}
\lref\rBrZi{E. Br\'ezin and J. Zinn-Justin, {\it Phys. Lett.} B288 (1992)
54.}
\lref\rHINS{S. Higuchi, C.Itoi, S. Nishigaki, N. Sakai, {\it Nucl. Phys.} B318 (1993) 63; {\it
Nonlinear Renormalization Group Equation for Matrix Models}}
\lref\rCorrel{B. Eynard, {\it Eigenvalue distribution of large random matrices,
from one matrix to several coupled matrices}, Nuclear Physics, to appear in 1997, {\it cond-mat}/9707005.}
\lref\rPZinn{P. Zinn-Justin, {\it cond-mat}/9703033 March 1997. {\it Random hermitian matrices in an external field}.}
\lref\rIZ{C. Itzykson, J.B. Zuber, {\it J. Math. Phys. }21 (1980) 411.}
\lref\rMehtaEyn{B. Eynard, M.L. Mehta, {\it Matrices coupled in a chain: eigenvalue correlations}. {\it Saclay preprint} SPhT 97/112, {\it cond-mat}/9710230. }

\def\tr{{\rm tr}\,}
\def\det{{\rm det}\,}

\def\ee#1{{\rm e}^{^{\textstyle#1}}}
\def\d#1{{\rm d}#1\,}

\def\Re{{\rm Re\,}}
\def\Im{{\rm Im\,}}

\def \l{\lambda}
\def \om{\omega}

\def\P{{\cal P}}

\def\t#1{\tilde{#1}}
\def\lra{\longrightarrow}

\def\encadremath#1{\vbox{\hrule\hbox{\vrule\kern8pt 
\vbox{\kern8pt \hbox{$\displaystyle #1$}\kern8pt} 
\kern8pt\vrule}\hrule}} \def\enca#1{\vbox{\hrule\hbox{
\vrule\kern8pt\vbox{\kern8pt \hbox{$\displaystyle #1$}
\kern8pt} \kern8pt\vrule}\hrule}}
\def\appendix#1{\vbox{\bigskip \noindent {\bf Appendix #1} \bigskip}}

\Title{\vbox{\baselineskip12pt\hbox{DTP 97-59, SPhT/98-001}}}
{\vbox{\hbox{\centerline{Correlation functions of eigenvalues of multi-matrix models,}}
\hbox{\centerline{\it and the limit of a time dependent matrix}}}}

\vskip-\smallskipamount
\centerline{B. Eynard\footnote{*}{e-mail: Bertrand.Eynard@durham.ac.uk}}

\smallskip{\baselineskip12pt
\centerline{Department of Mathematical Sciences}
\centerline{University of Durham, Science Laboratories}
\centerline{South Road, DURHAM DH1 3HP, U.K.}}

\medskip

\vskip.1in


\centerline{\bf Abstract:}
The universality of correlation functions of eigenvalues of large random matrices has been observed in various physical systems, and proved in some particular cases, as the hermitian one-matrix model with polynomial potential. 
Here, we consider the more difficult case of a unidimensional chain of matrices with first neighbour couplings and polynomial potentials.

An asymptotic expression of the orthogonal polynomials allows to find new results for the correlations of eigenvalues of different matrices of the chain.

Eventually, we consider the limit of the infinite chain of matrices, which can be interpreted as a time dependent one-matrix model, and give the correlation functions of eigenvalues at different times.




\smallskip
\Date{11/97, for Journal of Physics A}


It has now since a long time been observed experimentally or numerically  \refs{\rCG,\rHWB,\rCamarda} that the distribution of energy levels of disordered systems is universal in some regime.
For instance the connected correlation function of two levels separated by a small number of other levels does not depend on the details of the system, only on its symmetries, while the density of levels is very dependent of the specific details of the system.
It was thus conjectured that the correlation functions can be obtained from a gaussian random matrix model (the matrix might be the hamiltonian, the scattering or transfert matrix). Such a conjecture would be a kind of a central-limit-theorem for large random matrices.

This conjecture has been proved in the special case of a one hermitian matrix model with polynomial potential \refs{\rBrZe,\rWigner} and for a two-matrix-model \rCorrel .
It has also been noted that the connected correlation functions of more than two eigenvalues should present a stronger universality than the density itself.

The analysis of \refs{\rBrZe,\rCorrel} was based on the method of orthogonal
polynomials.
The correlation functions were expressed in terms of {\it kernels}, depending on two variables, which are given as sums of polynomials.
Those results are exact and have been known for a long time.
The problem was to find an asymptotic expansion in the large $N$ limit ($N$ is the size of the matrices), indeed those kernels involve the sum of $N$ polynomials of degree running from $0$ to $N-1$.
In the one-matrix model case \rBrZe , the Darboux-Christoffel theorem allows to rewrite the kernel with only two polynomials of degree $N$ and $N-1$.
Asymptotic expressions of orthogonal polynomials are then used to evaluate the universal correlation functions in the short range regime.

In \rCorrel\ it was claimed that this analysis could probably be extended to a more general case, which is the chain of $p$ random hermitian matrices $M_1,\dots M_p$, where each matrix $M_k$ is coupled linearly to the following one $M_{k+1}$.

In particular, when the number of matrices of the chain becomes infinite and the coupling are chosen appropriately, this model can be viewed as a time-dependent random matrix. The coupling between neighbouring matrices of the chain being then a kinetic term of the form $(\d{M}/\d{t})^2$.

Here, we will generalize the analysis of \refs{\rBrZe,\rCorrel} to the chain of matrices.
The paper is organized as follows:

\noindent The first section concerns the discrete chain, and the second section, the continuous-time limit.
In the first section, we first present the matrix-model, remind the orthogonal polynomial method, then we relate the correlation functions to the orthogonal polynomials via the kernels and generalize the Darboux Christoffel theorem in order to rewrite those kernels as a sum of a finite number of terms. A WKB approximation of the orthogonal polynomials allows to find asymptotic expressions of the kernels, and thus to find the correlation functions in the large $N$ limit.
We then conclude by examining the universal properties of those correlations.
%
%
\newsec{The Chain of matrices}
Let us first present the model and introduce the notations coherent with those of \rCorrel .
\par

Consider a linear chain of $p+1$ random hermitian $N\!\times\! N$ matrices  $M_i \, (0\leq i \leq p)$, with a probability:

\eqn\eProba{
{\cal P}(M_i)= {1\over Z}\, \prod_{i=0}^{p}\, \ee{-N\tr V_i(M_i)} \,\,\prod_{i=0}^{p-1}\,\ee{-Nc\,\tr (M_i -M_{i+1})^2} }
where the $V_i$ are polynomial potentials, $c$ is the coupling constant between neighbouring matrices, and $Z$ is the partition function.
( In the next section, we will consider the continuum limit of this model: the index $i$ will become a continuous variable: the time $t=i\epsilon$, and with $2c=1/\epsilon$, the linear term $\sum_i c\,\tr (M_i -M_{i+1})^2$ will become a kinetic term $\int {1\over2}\dot{M}^2(t)\, \d{t}$ ).

The Itzykson-Zuber formula \rIZ\ allows to integrate out the angular variables (the unitary group), and leaves us with the joint probability for the eigenvalues ( let us note $\l_{i;j}\, \, (0\leq j\leq N-1)$ the $j^{\rm th}$ eigenvalue of the matrix $M_i$):
\eqn\eProbval{
\rho(\l_{i;j})= {1\over Z} \Delta(\l_1)\Delta(\l_p) \prod_{i=0}^p \ee{-N\sum_j V_i(\l_{i;j}) }
 \,\prod_{i=0}^{p-1}\, {\rm det}_{k,l}\,
\left( \ee{-Nc(\l_{i;k}-\l_{i+1;l})^2}\right)
}
where the $\Delta$ are the Vandermonde determinants:
$$ \Delta(\l_i)= \prod_{k<l} \, (\l_{i;k}-\l_{i;l}) $$

We would now like to compute the conditional probabilities of some subset $I$ of these $N\!\times\! (p+1)$ eigenvalues.
We thus have to integrate \eProbval\ over all the eigenvalues which do not belong to $I$.
For instance, the density of the eigenvalues of $M_i$ is:
$$\rho_i(\l_{i;1})=\int \rho \prod_{(j,k)\neq (i,1)} \d{\l_{j;k}} ,$$
the correlation function of two eigenvalues of $M_i$ is:
$$ \rho_{ii}(\l_{i;1},\l_{i;2})= \int \rho \prod_{(j,k)\neq (i,1),(i,2)} \d{\l_{j;k}} $$
and the correlation function of two eigenvalues of two matrices $M_i$ and $M_j$ is:
$$ \rho_{ij}(\l_{i;1},\l_{j;1})= \int \rho \prod_{(l,k)\neq (i,1),(j,1)} \d{\l_{l;k}} $$

As in the one matrix case \refs{\rCorrel,\rBrZe}, all these densities and correlation functions can be calculated by the orthogonal polynomials method \rMehtaEyn , let us recall this method \rMehta .

\subsec{Orthogonal polynomials}

Consider two families of polynomials $\P_n$ and $\t{\P}_n$, of degree $n$, beginning with the same leading term, and which obey the orthogonality relation:
\eqn\edefortho{
\int \d{\l_0}\dots\d{\l_p} \, \ee{-N \sum_i V_i(\l_i)} \ee{-Nc\sum_i (\l_i-\l_{i+1})^2} \P_n(\l_0) \t{\P}_m(\l_p)=\delta_{n,m} }
we define the wave functions $\psi_n$ and $\t\psi_n$ by:
$$\psi_n(\l_0)=\P_n(\l_0)\,\ee{-N {1\over2}V_0(\l_0)}$$
$$\t\psi_n(\l_p)=\t{\P}_n(\l_p)\,\ee{-N {1\over2}V_p(\l_p)}$$
(note that the normalizations differ from \rCorrel ).
with the help of the orthogonality relation \edefortho\ , we can define two families of Hilbert Spaces $E_i,\t{E}_i$, and orthogonal functions in each of them:

\eqna\edefpsini
$$\eqalignno{
\psi_{0,n}(\l_0) = & \psi_n(\l_0) & \cr
\psi_{i;n}(\l_i)= & \int \d{\l_0}\dots\d{\l_{i-1}} \, \psi_n(\l_0)\,                    \ee{-Nc((\l_0-\l_1)^2+\dots+(\l_{i-1}-\l_i)^2)} \cr
 & \qquad \ee{-N({1\over2}V_0(\l_0)+V_1(\l_1)+\dots+V_{i-1}(\l_{i-1})+{1\over2} V_i(\l_i))} \edefpsini{}\cr
}$$
\eqna\edeftpsini
$$\eqalignno{
\t\psi_{p;n}(\l_p) = & \t\psi_n(\l_p) & \cr
\t\psi_{i;n}(\l_i) = & \int \d{\l_{i+1}}\dots\d{\l_p} \, \t\psi_n(\l_p)\,
 \ee{-Nc((\l_i-\l_{i+1})^2+\dots+(\l_{p-1}-\l_p)^2)}    \cr
  & \qquad \ee{-N({1\over 2} V_i(\l_i)+V_{i+1}(\l_{i+1})+\dots+V_{p-1}(\l_{p-1})+{1\over 2} V_p(\l_p))} \edeftpsini{}\cr
}$$
We shall note them with the convenient Dirac notations:
\eqn\edefbraket{
< n|_i=\psi_{i;n}  \qquad , \qquad |n>_i=\t\psi_{i;n} \,  .}
In the space $E_i$, we have the orthogonality relation:
$$\int \d{\l_i} \,\psi_{i;n}(\l_i)\t\psi_{i;m}(\l_i)=<n|m>_i=\delta_{n,m}$$
In each of these spaces, we can define the usual operators (acting on the right hand side: the ket):

 \noindent $\bullet \qquad\hat\l_i$, the operator which multiplies $\t\psi_i(\l_i)$ by $\l_i$.

 \noindent $\bullet \qquad\hat{P}_i={1\over N}{\partial \over \partial \l_i}$ which differentiate $\t\psi_i$ with respect to $\l_i$.
\medskip
These operators are defined only in the Hilbert space $E_i$. However all the $E_i$ are isomorphics, and an operator $\hat{O}$ initially defined in $E_i$  can also be defined in any $E_j$ by its matrix elements:
$$ <n|\hat{O} |m>_j \mathop{=}^{\rm def} <n|\hat{O} |m>_i =\int \d{\l_i} \, \psi_{i;n}(\l_i)\,{O}\, \t\psi_{i;m}(\l_i)
$$
From now on, we will drop the index $i$ for the bras and kets.

\subsec{Equations of motion}
 
From the former definitions we immediately obtain the equations of motion:
\eqn\ePi{ \hat{P}_i=2c(\hat\l_{i+1}-\hat\l_i)-{1\over2} V'_i(\hat\l_i) }
\eqn\ePiOM{ \hat{P}_0=2c(\hat\l_1-\hat\l_0)-{1\over2}V'_0(\hat\l_0) \qquad , \qquad \hat{P}_p=2c(\hat\l_p-\hat\l_{p-1}) +{1\over2}V'_p(\hat\l_p) }
and with an integration by parts:
\eqn\eMotionop{ \hat\l_{i-1}+\hat\l_{i+1}-2\hat\l_i={1\over 2c}V'_i(\hat\l_i) }

Let us now introduce more convenient notations.
Since we began with polynomials $\P_n$ and $\t{\P}_n$, we know how the multiplications or derivations by $\l_0$ or $\l_p$ will act:
multiplication by $\l_p$ raises the degree of $\t{\P}_n$ by 1, and $\l_p \t{\P}_n(\l_p)$ can be decomposed onto the base of the $\t{\P}_{n-k}$ with $k\geq -1$:
$$ \l_p \t{\P}_n (\l_p) = \alpha(n) \t{\P}_{n+1} (\l_p) + \sum_{k\geq 0} \t{\alpha}_k(n-k) \t{\P}_{n-k} (\l_p) $$
(where $\alpha(n)$ is the ratio of the leading coefficients of $\t{\P}_n$ and $\t{\P}_{n+1}$, and the $\t{\alpha}_k(n)$ are coefficients to be determined later).

Let us write this in operatorial notations.
In this purpose we introduce the operator $\hat{x}$ which decreases the level ( a kind of anihilation operator):
$$\hat{x}|n>=|n-1> \qquad , \qquad <n|\hat{x}=<n+1|$$
Although $\hat{x}$ is not invertible, we shall abusively write $\hat{x}^\dagger={1\over \hat{x}}$, for it will make no difference when we go to the large $N$ limit, and it will considerably simplify the notations \footnote{*}{we have $\hat{x}|0>=0$, and $\hat{x}^{-1}$ is not defined only on $|0>$. One solution could be to define a state $|-1>$, provided that all the $\alpha_k(-1)$ vanish, which is true. }.

We can then write:
\eqna\elopx
$$\eqalign{
\hat\l_0^\dagger = & {1\over\hat{x}}\alpha(\hat{n})+\sum_{k\geq 0} \hat{x}^k \alpha_k(\hat{n})\cr
\hat\l_p = & {1\over\hat{x}}\alpha(\hat{n})+\sum_{k\geq 0} \hat{x}^k \t\alpha_k(\hat{n})}\hskip 4cm \elopx{}$$
Remember that $\hat{\l}_0$ acts on the ket $|n>$, i.e. on the polynomial $\t{\P}_n$, its adjoint acts on the bra $\P_n$.
Note also that the first term, $\alpha(n)$ is the same for both $\l_0$ and $\l_p$ because we have chosen the polynomials $\P_n$ and $\t{\P}_n$ with the same leading coefficient.

Similarly, noting that the operator $\hat{P}={1\over N}{\d\over \d{\l_p}}$ decreases the degree of the polynomial $\t{\P}_n(\l_p)$ by 1:
$$ {\d \over \d{\l_p}} \t{\P}_n(\l_p) = {1\over \alpha(n-1)}\, n \t{\P}_{n-1} + \ldots $$
we can can express the operators $\hat{P}_0$ and $\hat{P}_p$ in power series of $\hat{x}$:
\eqn\eMotioniniop{ \hat{P}^\dagger_0+{1\over 2}V'_0(\hat\l^\dagger_0)= {1\over \alpha(\hat{n})}\hat{x}{\hat{n}\over N} + O(\hat{x}^2) }
\eqn\eMotioniniopb{ \hat{P}_p+{1\over2}V'_p(\hat\l_p) = {1\over \alpha(\hat{n})}\hat{x}{\hat{n}\over N} + O(\hat{x}^2) }

\medskip
We might as well write any of the operators $\hat\l_i$ with such notations:
$$ \hat\l_i = \sum_k   \alpha_{i;k}(\hat{n}) \hat{x}^k ,$$
but let us first go to the large $N$ limit.

\subsec{large $N$ limit}

In the classical limit $N\to\infty$, all those operators become numbers. Indeed,  the commutators $[\hat{P},\hat\l]=1/N$ and $[\hat{x},\hat{n}/N]=\hat{x}/N$ are proportional to $1/N$ which thus plays the role of $\hbar$ \footnote{**}{
actually, this is true only if the support of the density is connected, i.e. we assume we have a one-cut solution, for a $k$-cut solution, we would need to consider the operators as $k\times k$ matrices.
e.g. for a symetric double well, one needs to distinguish between even and odd values of $n$, which introduces two sets of coefficients $\alpha_k(2n)$ and $\alpha_k(2n+1)$.}.
We then write:
\eqn\edegalpha{\l_i(x)=\sum_k \alpha_{i,k} x^k \qquad {\rm with}\qquad -\prod_{j=i+1}^{p}\deg V'_{j}\,\leq k \leq \, \prod_{j=0}^{i-1} \deg V'_{j} }
The bounds on $k$ are easily derived from the equations of motion and boundary conditions.
We also consider the limit where $n$ is large and close to $N$, so that the $\alpha_{i,k}$ don't depend anymore on $n$, they are just coefficients.

In addition, There exist a remarkable relation (the proof from the cannonical commutation relations is not difficult but of no interest for what follows):
\eqn\ecomrule{ 1 = c \sum_k k(\alpha_{i+1,k}-\alpha_{i-1,k})\alpha_{i,-k} .}
Let us rewrite in the classical limit the equations of motion \eMotionop\ and boundary conditions \eMotioniniop \eMotioniniopb\ previously written for operators.
We have the following system of equations:
\eqn\emotion{ \l_{i-1}+\l_{i+1}-2\l_i={1\over 2c}V'_i(\l_i) \qquad 1\leq i \leq p-1 }
with the boundary conditions:
\eqn\einiO{ V'_0(\l_0)-2c(\l_1-\l_0) = {1\over \alpha x} + O(1/x^2) }
\eqn\einiM{ V'_p(\l_p)-2c(\l_{p-1}-\l_p) = {x\over \alpha} + O(x^2) }
One can verify that we have exactly as many equations as unknowns. 
If we were able to solve this system of algebraic equations and determine all the $\alpha_{i,k}$, we could define $p+1$ functions $\l_i(x)$, of an auxillary variable $x$.
We will see below the important role they play.

\subsec{WKB approximation}

One can find (by a simple generalization of \rCorrel , i.e. by a kind of saddle point method for matrix integrals to the explicit expressions given in appendix B) some asymptotic expressions of the $\psi_{i;n}$ in the limit $N$ large and $|n-N|\sim O(1)$:
\eqna\eWKB
$$\eqalign{
\psi_{0;n}(\l_0)&\sim \sum_{x \, / \, \l_0(x)=\l_0} \,\left({\pi\over Nc}\right)^{-p/4}{1\over \sqrt{2\pi}}{1\over\sqrt{\l'_0(x)}} x^{n-N} \ee{-2Nc\int^x (\l_1-\l_0)\l'_0}\ee{{N\over2} V_0(\l_0)}\cr
\psi_{i;n}(\l_i)&\sim \sum_x \, \left({\pi\over Nc}\right)^{i/2-p/4}{1\over \sqrt{2\pi}}{1\over\sqrt{\l'_i(x)}} x^{n-N} \ee{-Nc\int^x (\l_{i+1}-\l_{i-1}) \l'_i}\cr
\psi_{p;n}(\l_p)&\sim \sum_x \, \left({\pi\over Nc}\right)^{p/4}{1\over \sqrt{2\pi}}{1\over\sqrt{\l'_p(x)}} x^{n-N} \ee{-2Nc\int^x (\l_p-\l_{p-1}) \l'_p}\ee{-{N\over2} V_p(\l_p)}\cr
\t\psi_{0;n}(\l_0)&\sim \sum_x \, \left({\pi\over Nc}\right)^{p/4}{1\over \sqrt{2\pi}}{1\over\sqrt{-\l'_0(x)}} x^{N-n-1} \ee{2Nc\int^x (\l_1-\l_0) \l'_0} \ee{-{N\over2} V_0(\l_0)}\cr
\t\psi_{i;n}(\l_i)&\sim \sum_x \, \left({\pi\over Nc}\right)^{p/4-i/2}{1\over \sqrt{2\pi}}{1\over\sqrt{-\l'_i(x)}} x^{N-n-1} \ee{Nc\int^x (\l_{i+1}-\l_{i-1}) \l'_i}\cr
\t\psi_{p;n}(\l_p)&\sim \sum_x \, \left({\pi\over Nc}\right)^{-p/4}{1\over \sqrt{2\pi}}{1\over\sqrt{-\l'_p(x)}} x^{N-n-1} \ee{2Nc\int^x (\l_{p}-\l_{p-1}) \l'_p}\ee{{N\over2} V_p(\l_p)}\cr}
\eWKB{}$$
We shall not prove those asymptotic expressions, but just give some intuitive explanations.

\noindent - First, observe that at leading order, all of them have the form:
$$\psi_i \sim \ee{-N\int^{\l_i} P_i \, \d{\l_i} } \qquad {\rm and} \qquad \t\psi_i \sim \ee{N\int^{\l_i} P_i \, \d{\l_i} } $$
which is simply the solution of the differential equation $ \hat{P}_i |n> = {1\over N} {\d\over \d{\l_i}} |n>$ in the large $N$ limit .
The lower bound of the integral, which has not been written here for simplicity, depends on $i$, it is determined by the condition that $\int^{\l_i} V'_i(\mu_i)\d{\mu_i} = V_i(\l_i) $ .
 
\noindent - The $x^{n-N}$ term comes from the definition of $\hat{x}$:
$$ <N| \hat{x}^{n-N} = <n|  \, $$

\noindent - Moreover, observe that the approximation for $\psi_{i+1}$ can be derived from $\psi_i$ by steepest descent in \edefpsini\ , and the expressions for the $\t\psi$'s can be derived from the $\psi$'s by $x\to 1/x$ and $i\to p-i$.

\noindent - Finally, the normalization constants and the $1/\sqrt{2\pi\l_i'(x)}$ are just what is needed to satisfy the normalization condition
$$<n|m>=\delta_{nm} \, .$$
Remark that all that is nothing else than WKB approximation.\par\medskip
Remember that, in quantum mechanics, the wave function of a particle outside a potential well decreases exponentially, while inside the well it is a stationary wave, i.e. a superposition of two opposite progressing waves.
This is also what we have here:

\noindent - the sum over $x$ means that you have to consider the values of $x$, solutions of $\l_i(x)=\l_i$ which have this property.
When $\l_i$ belongs to $[a_i,b_i]$ (the suport of the density of eigenvalues of the $i^{\rm th}$ matrix), the equation $\l_i(x)=\l_i$ has no real solution, it has only pairs of complex conjugate solutions, which give the stationary wave.
The sum of the two complex solutions will give rise to some real expression for $\psi_{i;n}$, involving cosinus and sinus functions instead of exponentials (cf \refs{\rBrZe,\rCorrel}).
When $\l_i$ is outside $[a_i,b_i]$, you have to keep only the solution which decreases exponentially at infinty.

From now on, we will consider only the first case, i.e. $\l_i \in [a_i,b_i]$.

\subsec{Kernels}

Remember that we have introduced the orthogonal polynomials in order to integrate the joint-density \eProbval\ over a subset $I$ of the $N\times (p+1)$ variables.
In this purpose, let us as  usual \rMehta\ rewrite the Vandermonde determinants:
$$ \Delta(\l_0) = \prod_{k<l} (\l_{0;k}-\l_{0;l}) = \mathop{\det}_{k,l} \left( \l_{0;l} \right)^k  $$
Since linear combinations of columns preserve the determinant, we can rewrite:
$$\Delta(\l_0)={\rm cte}\,\mathop{\det}_{k,l} \P_k(\l_{0;l})$$
$$\Delta(\l_p)={\rm cte}\,\mathop{\det}_{k,l} \t{\P}_k(\l_{p;l})$$
The ${\it \rm cte}$ is a normalization which comes from the fact that the polynomials $\P_k$ and $\t{\P}_k$ are not monic (actually ${\it \rm cte}=\prod_{n=0}^{N-1} \alpha(n)^{N-1-n}$).
Any partial integration of \eProbval\ can thus be written as an integral over the $\psi_{i;n}$ and $\t\psi_{j;m}$. Since they are orthogonal, the integration is easily performed, and the final result can be written in terms of $2(p+1)^2$ Kernels defined by:
\eqn\edefK{K_{i,j}(\l_i,\l_j)={1\over N}\sum_{n=0}^{N-1} \psi_{i;n}(\l_i)\t\psi_{j;n}(\l_j)}
and
\eqn\edefE{
E_{i,j}(\l_i,\l_j)=
\left\{
\eqalign{
 & 0  \cr
& \ee{-{N\over 2}\left[ 2c(\l_i-\l_{i+1})^2 + V_i(\l_i)+V_{i+1}(\l_{i+1})\right]} \cr
& \int \prod_{l=i+1}^{j-1} \d{\l_l} \, \prod_{l=i}^{j-1} \, E_{l,l+1}(\l_l,\l_{l+1}) \cr
}
\quad \eqalign{ &  {\rm if}\quad i\geq j \cr &  {\rm if}\quad  i=j-1 \cr & \cr &   {\rm if}\quad i< j-1 \cr}
\right. }
In the $p=1$ case discussed in \rCorrel\ there were only four kernels (the $K_{ij}$), indeed, the $E_{ij}$ which were just numbers were absorbed into the normalizations.
But in the general case, the $E_{ij}$ contain integrations and cannot be absorbed.
Note that the $E_{ij}$ are the propagators from $\psi_i$ to $\psi_j$ ($i<j$):
$$ \int \d{\l_j} \t\psi_{j;n}(\l_j) E_{i,j}(\l_i,\l_j)= \t\psi_{i;n}(\l_i) $$
$$ \int \d{\l_i} \psi_{i;n}(\l_j) E_{i,j}(\l_i,\l_j)= \psi_{j;n}(\l_j) $$

We thus have the following projection relations:
\eqna\eprojEK
$$\eqalign{
 \int \d{\l_j} E_{ij} E_{jl} = & E_{il} \qquad {\rm if} \quad i< j< l \cr
 \int \d{\l_j} E_{ij} K_{lj} = & K_{li} \qquad {\rm if} \quad i< j \cr
 \int \d{\l_i} E_{ij} K_{il} = & K_{jl} \qquad {\rm if} \quad i< j \cr
 \int \d{\l_j} K_{ij} K_{jl} = & {1\over N} K_{il} \cr
} \qquad\eprojEK{}$$

\subsec{Correlation functions}

In terms of these kernels, the joint density \eProbval\ of all the eigenvalues of all the matrices can be rewritten:
$$ \rho= {\rm cte}\,\,\,\det K_{0,p}\, \det E_{0,1} \det E_{1,2} \dots \det E_{p-1,p} $$
To obtain the densities and correlation functions of some set of eigenvalues, we have to partially integrate $\rho$ with respect to the other eigenvalues, and this can be done \rMehtaEyn\ with the help of the projection rules \eprojEK{} .
The general result is given in appendix A.
Here, we will only consider the one and two-points functions.

The density of eigenvalues (the one-point function) of the $i^{\rm th}$ matrix is:
\eqn\erhoun{ \rho_i(\l_i)=K_{i,i}(\l_i,\l_i)}
and the two point connected correlation function of one eigenvalue $\l_i$ of the $i^{\rm th}$ matrix , and one eigenvalue $\mu_j$ of the $j^{\rm th}$ matrix is:
\eqn\erhocdeu{ \rho^{(c)}_{i,j}(\l_i,\mu_j)= -K_{i,j}(\l_i,\mu_j)K_{j,i}(\mu_j,\l_i)+{1\over N}K_{i,j}(\l_i,\mu_j)E_{i,j} (\l_i,\mu_j) \qquad (i\leq j) }

We now have to evaluate the kernels $K_{ij}$ and $E_{ij}$ in the large $N$ limit.
The first step will be a generalization of the Darboux-Christoffel theorem, which allows one to rewrite $K_{ij}$ as a sum of a small number of terms, instead of the sum of $N$ terms as in \edefK\ .
The second step will be to use the WKB approximations \eWKB\ ~for the $\psi$'s.
The kernels $E_{ij}$ will be evaluated by steepest descent.

\subsec{Generalization of the Darboux-Christoffel theorem for the kernels}

As in \rCorrel\ the Darboux-Christoffel theorem can be generalized.
Formally, we write that
\eqn\eKxy{ \psi_{i;n}(\l_0) = \hat{x}^{-(N-n)} \psi_{i;N}(\l_0) \qquad , \qquad \t\psi_{j;n}(\l_p) = \hat{x}^{N-n} \t\psi_{j;N}(\l_p) }
and we sum up the geometrical series in \edefK\ :
\eqn\eDarbou{ K_{i,j}(\l_i,\l_j) = {1\over N}\,{\hat{y}\over \hat{x}-\hat{y}} \, \psi_{i;N}(\l_i(x))\,\t\psi_{j;N}(\l_j(y))) }
(we have called $\hat{y}$ the operator $\hat{x}$ acting on the second variable).
Multiplying both sides of \eDarbou\ by $(\l_i(\hat{x})-\l_i(\hat{y}))$ would give in the left hand side a differential polynomial acting on $K_{ij}$ (indeed $\l_i(\hat{y})$ can be rewritten as a polynomial in $\hat{\l_j}$ and $\hat{P}_j$ with the help of eq.\ePi ,\eMotionop ), and in the right hand side a polynomial in $\hat{x}$ and $\hat{y}$, i.e. a small number of $\psi_{i;n}$ and $\t\psi_{j;n}$ with $|n-N|\ll N$, 
(an explicit example is given in Appendix C).
However we shall not do it, but use directly \eDarbou\ in the large $N$ limit, where $\hat{x}$ and $\hat{y}$ become numbers $x$ and $y$.

The kernels can thus be approximated by:
$$K_{i,j}(\l_i,\l_j) \sim {1\over N}\,{y\over x-y}\,\psi_{i;N}(\l_i(x))\,\t\psi_{j;N}(\l_j(y))$$
and using the WKB asymptotic expressions \eWKB{}\ of $\psi_{i;N}$ and $\t\psi_{j;N}$:
\eqn\eKNgd{K_{i,j}\sim \sum_{x y} \left({\pi\over Nc}\right)^{i-j\over 2}{1\over 2i\pi N}\, {1\over x-y} \,{1\over \sqrt{\l'_i(x)\l'_j(y)}}
\,\ee{-Nc\left( \int^x (\l_{i+1}-\l_{i-1})\l'_i\, -\int^y (\l_{j+1}-\l_{j-1})\l'_j \right)} }
where as usually, the $x$ and $y$ stand for the complex solutions of $\l_i(x)=\l_i$ and $\l_j(y)=\l_j$.

One can also find an asymptotic expression for the kernel $E_{ij}$ by steepest descent:
\eqn\eENgd{ E_{ij}\sim \left({\pi\over Nc}\right)^{j-i-1\over2} {1\over \sqrt{ D_{ij}(x,y)\,}} \,\ee{-Nc\, U_{ij}(\l_i,\l_j)} }
where
$$U_{ij}(\l,\mu)=\sum_{l=i}^{j-1}\,\left( (\l_l-\l_{l+1})^2 + {1\over2c}V_l(\l_l)+{1\over2c}V_{l+1}(\l_{l+1}) \right) $$
where the $\l_l(\l,\mu)$ are determined by the saddle point equation:
$$ \l_i=\l \quad , \quad \l_j=\mu \qquad , \quad 2\l_l+{1\over 2c}V'_l(\l_l) = \l_{l-1}+\l_{l+1} \quad {\rm for} \,\, i<l<j $$
and $D_{ij}$ is the determinant of the matrix of second derivatives of $U_{ij}$ with respect to the $\l_l$'s:
$$ D_{ij} = \det \left| {\matrix{ 2+{V''_{i+1}(\l_{i+1})\over 2c} & -1 & 0 & \dots & 0 & 0 \cr
                                  -1 & 2+{V''_{i+2}(\l_{i+2})\over 2c} & -1 & 0 & \dots  & 0 \cr
                                 0  & \ddots & \ddots & \ddots & & \vdots \cr
                                   & & & & & \cr
                                   \vdots &   & 0 & -1 & 2+{V''_{j-2}(\l_{j-2})\over 2c} & -1 \cr
                                  0 & 0 & \dots & 0 & -1 & 2+{V''_{j-1}(\l_{j-1})\over 2c} \cr }}   \right|  \,\, .$$

In the particular case $x=y$ we have:
$$ U_{ij}(x,x) = -\int^x (\l_{i+1}-\l_{i-1})\l'_i\, +\int^y (\l_{j+1}-\l_{j-1})\l'_j \,\, ,$$
$$ {\rm and} \qquad D_{ij}(x,x) = \sum_{l=i}^{j-1} {\l'_i(x)\l'_j(x)\over \l'_l(x) \l'_{l+1}(x)} \,\, .$$

Substituting \eKNgd\ and \eENgd\ into \erhoun\ and \erhocdeu\ we can now evaluate the correlation funtions.

\subsec{Correlation functions in the short distance limit}

\noindent $\bullet$ case $i=j$

Setting $i=j$ in \eKNgd\ gives:
$$K_{i,i}(\l,\mu)\sim \sum_{xy} \, {1\over 2i\pi N} {1\over x-y} {1\over \sqrt{\l'_i(x)\l'_i(y)}} \,\ee{-Nc \int_y^x (\l_{i+1}-\l_{i-1})\l'_i}  $$
where $x$ and $y$ are the complex solutions of $\l_i(x)=\l$ and $\l_i(y)=\mu$.
When $\l$ is close to $\mu$, at leading order, we keep only the values of $x$ and $y$ such that $|x-y|$ is small, then this reduces to:
$$K_{i,i}(\l,\mu)\sim - {1\over \pi N} {1\over \l-\mu} \,\sin{\left\{ Nc(\l-\mu) {\rm Im}(\l_{i+1}-\l_{i-1}) \right\} } \,\,\ee{-Nc (\l-\mu){\rm Re} (\l_{i+1}-\l_{i-1}) }  $$

In particular, when $\l=\mu$ we obtain the density:
\eqn\erhoununiv{\encadremath{\rho_i(\l)=K_{i,i}(\l,\l)=-{c\over \pi}\Im (\l_{i+1}(x)-\l_{i-1}(x)) =-{1\over \pi}\Im P_i(x) } }

When $\l$ is close to $\mu$ but different, we can compute the two-point connected correlation function:
$$\rho^{(c)}_{i,i}(\l,\mu)= - K_{i,i}(\l,\mu) K_{i,i}(\mu,\l)$$
i.e.
\eqn\erhodeuuniv{\encadremath{\rho^{(c)}_{i,i}(\l,\mu)\mathop\sim_{\l\to\mu}  -\left( {\sin{N\pi(\l-\mu)\rho_i(\l)}\over N\pi(\l-\mu)} \right)^2 } }
we recover the universal two point correlation function in the short distance regime.

\medskip

\noindent $\bullet$ case $i\neq j$

It is now meaningless to consider the limit $\l_i$ close to $\l_j$ since they  are eigenvalues of different matrices.
Generically, $K_{ij}$ is of order $1/N$, which means that the connected correlation function is of order $1/N^2$, and we can say that in the large $N$ limit $\l_i$ and $\l_j$ are uncorrelated.

The only limit in which the correlation may become larger than $1/N^2$ is the case where $x-y$ becomes small.
The equation $x=y$ defines a function $\l_i(\l_j)$.
The problem is that this function takes complex values in the interesting domain - for example we see from eq\erhoununiv\ that $\l_{i+1}(\l_i)$ or $\l_{i-1}(\l_i)$ take complex values, ( this fact has already been debated in \rCorrel\  ) - and we have not found any physical interpretation, except in the case of the continuous model described in the next section.

However, let us assume that $|x-y|$ is small (we will also assume $i<j$).
We introduce the scaling variable:
$$ \Delta= \sqrt{N} \,(x-y)\,\sqrt{\l'_i \l'_j} \sqrt{c\over D_{ij}} $$
In the limit $|x-y|\to 0$, the Taylor expansion of the term appearing in the exponential in \eKNgd\ gives:
$$ \int^x (\l_{i+1}-\l_{i-1})\l'_i\, -\int^y (\l_{j+1}-\l_{j-1})\l'_j = - U_{ij} + {\l'_i\l'_j\over D_{ij}}(x-y)^2 + O((x-y)^3)$$
Therefore, we have:
\eqna\eKExx
$$ \eqalign{
 K_{ij}\sim & \left({\pi\over Nc}\right)^{i-j\over2} {1\over 2i\pi \sqrt{N}} \sqrt{c\over D_{ij}}\, \ee{+Nc U_{ij}} \,\, {1\over \Delta} \ee{- \Delta^2} \cr
 K_{ji}\sim & \left({\pi\over Nc}\right)^{j-i\over2} {1\over 2i\pi \sqrt{N}} \sqrt{c\over D_{ij}}\, \ee{-Nc U_{ij}} \,\, {1\over \Delta} \ee{ \Delta^2} \cr
 {1\over N}E_{ij}\sim & \left({\pi\over Nc}\right)^{j-i\over2} {1\over \sqrt\pi\sqrt{N}} \sqrt{c\over D_{ij}}\, \ee{-Nc U_{ij}}   \cr
} \qquad\eKExx{}$$
Remember that each of these expressions is actually a sum over the different values of $x$ and $y$ satisfying $\l_i(x)=\l$ and $\l_j(y)=\mu$.
But since $\l$ and $\mu$ are not assumed real, the different values of $x$ or $y$ contributing to this sum are not the complex conjugates of each other, and the exponential term cannot simply be replaced by a sinus function.

\noindent However, we have the correlation function:
\eqn\erhodeushort{ \encadremath{ \rho^{(c)}_{i,j} \mathop\sim_{\Delta\to 0}
\sum_{xy} -{1\over 4\pi^2 N}  {c\over D_{ij}} {1\over\Delta}\ee{NcU_{ij}}
\sum_{xy} \left({1\over\Delta}-2\sqrt\pi \ee{-\Delta^2} \right)\ee{-NcU_{ij}} }}

To go further one would have to make some asumption on the argument of $x$ and $y$ (i.e. where in the complex plane are $\l$ and $\mu$), and then separate the imaginary and real parts of $U_{ij}$ and $\Delta$.
We would then observe that in this scaling regime ($|x-y|\sim 1/\sqrt{N}$),   $\rho_{i,j}^{(c)}$ is of order $O(1/N)$ instead of $O(1/N^2)$, and that it presents some kind of universality: it doesn't depend very explicitely on the potentials $V_i(\l)$.
The signification of this correlation function is therefore not clear, and has to be further studied.
However, this calculation was performed to prepare the next section, the continuous chain of matrices, where it is possible to have both $\l$ and $\mu$ real and $(x-y)$ small, because $D_{ij}$ will be small too.

\subsec{Smoothed functions}
If we smooth expression \erhocdeu , (we keep in the sums over $x$ and $y$ only the terms which don't oscillate at frequency $N$, i.e. for which the exponentials exactly cancel), we obtain:
\eqn\erhoijsmooth{ \rho^{(c)}_{i,j}(\l_i,\l_j)_{\rm smooth}\,\sim\, -{1\over 4\pi^2 N^2}{x'(\l_i) y'(\l_j)\over (x(\l_i)-y(\l_j))^2}   + \,{\rm complex \, conjugate} }
This result is exactly the same as for two matrices \refs{\rDKK,\rCorrel}~.


\newsec{continuum limit}

The model of a chain of matrices is naturally extended to the model of a time dependent random matrix.
We replace the integer index $i$ by a continuous time $t=i\epsilon$, wich runs from $0$ to $T=p\epsilon$.
The coupling term $c(M_i-M_{i+1})^2$ becomes a kinetic term, we set:
$$c={1\over 2\epsilon}$$
$$\sum_i \longrightarrow {1\over \epsilon}\int \d{t}$$
$$V_i(\l) \longrightarrow \epsilon V(\l,t)$$
The partition function thus becomes a functionnal integral for a one dimensional matrix field:
$$ Z=\int D[M(t)]\,\, \ee{-N\tr \int_0^T \d{t} \left({1\over 2}\left( {\d{M}\over\d{t}}\right)^2 + V(M(t),t)\right) }$$

We will now just make use of the results of the previous section for the discrete time case, and translate them into the language of the continuous model.

\medskip
\noindent We now have a set of orthogonal wave functions depending on time
$$\psi_{i;n}(\l)\longrightarrow \psi_n(\l,t)  \qquad {\rm and} \qquad \t\psi_{i;n}(\l)\longrightarrow \t\psi_n(\l,t)$$
which satisfy the orthogonality relationship:
$$ \int \d\l  \psi_n(\l,t) \t\psi_m(\l,t) = \delta_{n,m} $$
Actually, one has to change the normalization, just to absorb the constant factor $ (\pi/Nc)^{i/2} $ of eq \eWKB{}\ which becomes infinite in the continuum limit.
But this doesn't change anything else.

\subsec{The function $\l(x,t)$}

We can define a family of functions $\l(x,t)$ of an auxilary variable $x$:
\eqn\edeflxt{
\l_i(x)\lra \l(x,t)=\sum_k \alpha_k(t) x^k
}
Note that acording to eq \edegalpha , $k$ runs from $-\infty$ to $+\infty$ (except for the gaussian case), so that \edeflxt\ has to be taken as a formal expansion in Laurent Series.

\noindent $\l(x,t)$ obeys the continuous limit of the equations of motion \emotion :
\eqn\emotioncont{
{{\rm \partial}^2\over \partial{t^2}}\l(x,t)=V'(\l(x,t))
}
with the boundary conitions \einiO :
\eqna\eboundcont
$$\eqalign{
 \l(x,0)    = \alpha x+ \sum_{k\leq 0} \alpha_k(0) x^{-k}   \qquad & ; \qquad    \dot\l(x,0)=-{1\over x\alpha}+O(1/x^2)    \cr
 \l(x,T)    = {1\over x} \alpha + \sum_{k\geq 0} x^k \alpha_k(T) \qquad & ; \qquad \dot\l(x,T)={x\over \alpha}+O(x^2) \cr
} \qquad\qquad\eboundcont{} $$
those last conditions can be rewritten in a way which doesn't involve any expansion in powers of $x$ \footnote{***}{Actually, even in the discret case, $\omega_0(\l_0(x))=2c(\l_0(x)-\l_1(x))+V'_0(\l_0(x))$ is the resolvant of the first matrix  $\omega_0(\l)={1\over N}<\Tr {1\over \l-M_0}>$, and $\omega_p(\l_p(x))=2c(\l_p(x)-\l_{p-1}(x))+V'_p(\l_p(x))$ is the resolvant of the $p^{\rm th}$ matrix. 
In the continuum limit we have $\om(\l(x,0),0)=-\dot\l(x,0)$ and $\omega(\l(x,T),T)=\dot\l(x,T)$.
For the intermediate values of $i$ (intermediate times), there is no simple relationship between the resolvant $\om_i(\l)$ ($\om(\l,t)$) and the functions $\l_i(x)$ ($\l(x,t)$). }:
$$ \dot\l(x,0) \sim -{1\over \l(x,0)}\qquad {\rm when} \quad \l(x,0)\to\infty $$
$$ \dot\l(x,T) \sim {1\over \l(x,T)} \qquad {\rm when} \quad \l(x,T)\to\infty $$
however, to define the function $x(\l,t)$ you can't avoid performing the formal expansion.

\medskip
\noindent {\it Remark:} eq \ecomrule\ becomes:
$\sum_k k\dot\alpha_k \alpha_{-k} =1$

\subsec{The momenta}
The momentum $P(x,t)$ is the time derivative of $\l(x,t)$ (eq \ePi\ and \ePiOM ):
$$P(t)=\dot\l=\left.{\partial \l\over \partial t}\right|_{x}=v(\l,t)-i\pi\rho(\l,t)$$
its imaginary part $\rho(\l,t)={-1\over\pi}\Im P(t)$ is the density of eigenvalue at time $t$.

\subsec{The Kernels}

The discrete kernels $K_{ij}$ and $E_{ij}$ naturally have some continuous version:
$$K_{i,j}(\l,\mu)\lra K(\l,t | \mu,t')={1\over N}\sum_{n=0}^{N-1} \psi_n(\l,t)\t\psi_n(\mu,t')$$
$$E_{i,j}(\l,\mu)\lra E(\l,t | \mu,t')=\int_{\l(t)=\l}^{\l(t')=\mu}\, D[\l(\tau)]\,\, \ee{-N\int_t^{t'} \d\tau  \left( {1\over 2}{\dot\l}^2 + V(\l(\tau),\tau) \right)}$$
$E$ is the usual quantum mechanics propagator for a single particle in the potential $V$:
$$ \psi_n(\l,t')=\int \d\mu \psi_n(\mu,t)\, E(\mu,t | \l,t')$$
And those Kernels allow to calculate the two-point connected correlation function ($t<t'$):
$$ \rho^{(c)}(\l,t |\mu,t') = -K(\l,t | \mu,t') \left( K(\mu,t' | \l,t) -{1\over N} E(\l,t | \mu,t') \right) $$

\subsec{Correlations}

Let us calculate the 2-point correlation in the limit where $\delta\l=\l-\mu$  and $\delta t=t-t'$ are small, of order $1/N$.
In this purpose, we have to evaluate the kernel $K(\l,t | \l+\delta\l,t+\delta t)$ with the WKB approximation \eKNgd\ (remember that we have absorbed the $(\pi/Nc)$ factors):
$$ K \sim 2\Re{1\over 2i\pi N}{1\over \sqrt{\l'\l'}}{-1\over \delta x}\, \ee{-N\left( \int^\l P(\l',t) \d{\l'} - \int^{\l+\delta\l} P(\l',t+\delta t)\d{\l'} \right)}$$
where $P(\l,t)=\dot\l=\left({\partial \l / \partial t}\right)|_x = v-i\pi\rho$. 
Writting $\delta \l=\l'\delta x + P \delta t$ the denominator $\l'\delta x$ is $ \delta\l -v\delta t + i\pi\rho\delta t$, and the term in the exponential is $P\delta\l + \delta t \int^\l \left({\partial P/\partial t}\right)|_\l$ which is evaluated through a play with the partial derivatives at $\l$ fixed or $x$ fixed:
$$ V'(\l) = \ddot\l= \left.{\partial P\over \partial t}\right|_x = \left.{\partial P\over \partial t}\right|_\l + \left.{\partial P\over \partial \l}\right|_t\left.{\partial \l\over\partial t}\right|_x $$
We thus obtain:
\eqn\eKcontNgd{
 K\sim \Im { 1 \over N\pi(v\delta t-\delta\l-i\pi\rho\delta t)}
\ee{Ni\pi\rho(v\delta t-\delta\l)  }\,\ee{N(v\delta\l+\delta t(V-{1\over2}v^2+{1\over2}\pi^2\rho^2))}
}
Similarly, observing that the continuous limit of $D_{ij}$ is
$$ {1\over c}D_{ij} \sim 2\delta t $$
the asymptotic expression \eENgd\ for $E$ turns into:
\eqn\eEcontNgd{
{1\over N}E\sim {1\over \sqrt{2 \pi N \delta t}} \,  \ee{-N ( {1\over 2}{\delta\l^2\over\delta t} +V \delta t  ) } \qquad {\rm for} \quad \delta t>0 
}

\noindent $\bullet$ equal time correlations:

When $\delta t=0$ we recover the usual one-matrix-model result:
$$ K \sim {\sin{(N\pi\rho\delta\l)}\over N\pi \delta\l} \ee{N v\delta \l}$$
For instance when $\delta\l=0$ we obtain the one-point function (the density)  $K=\rho$.
And when $\delta\l\neq 0$ but of order $1/N$, we have the universal correlation function:
$$\encadremath{ \rho^{(c)}(\l,t \, |\, \l+\delta\l,t) \sim  -\rho^2 \, \left( {\sin{(N\pi\rho\,\delta\l)}\over N\pi\rho\delta\l} \right)^2 }$$
So, at equal time, the correlation function of close eigenvalues is still universal.

\medskip
\noindent $\bullet$ different times:

When $\delta t\neq 0$, the denominator of $K$ never vanishes.
It reaches its minimum for $\delta \l = v\delta t$.
We shall thus consider the regime:
$$ \delta \l = v\delta t \qquad {\rm and } \qquad \delta t \sim 1/N$$
where we have:
$$ K\sim {\rho\over N \pi^2 \rho^2 \delta t} \,\ee{N\delta t(V+{1\over2}v^2+{1\over2}\pi^2\rho^2)}$$
and
$$ {1\over N}E\sim {\rho\sqrt\pi \over \sqrt{2 N \pi^2 \rho^2 \delta t}} \, \ee{-N\delta t(V+{1\over2}v^2)}$$
Thus
\eqn\erhodift{ \encadremath{\rho^{(c)}(\l,t \, |\, \l+v\delta t,t+\delta t) \sim  -\rho^2 \,{1\over \Delta} \left( {1\over \Delta } + \sqrt{\pi\over 2}\,{\ee{{1\over2}\Delta} \over \sqrt{ \Delta}} \right)  \qquad {\rm with}\quad \Delta= N\pi^2\rho^2\delta t}}
which is remarkably universal, it doesn't depend on the precise form of the potential $V(\l,t)$.

\newsec{Conclusion}

We have thus, as announced in \rCorrel , generalized the properties of the correlation functions of the two matrix model to the chain of matrices.

As expected, the chain of matrices obeys the same universal properties as the one and two-matrix models.
We obtain similar results:

\noindent - the correlation function is generically of order $1/N^2$, except in the regime where the denominator of the kernel $K$ becomes small.
This may happen only at small distance $\delta\l$ and at small time interval $\delta t$.

\noindent - At equal time $(\delta t=0)$ we recover exactly the well known universal correlation function of the one-matrix model.
This is therefore one more matrix model where this universality is proved.

\noindent - At different times ($\delta t>0$), the 2-point connected correlation function is again a universal function of $\delta t$ and the shifted interval $\delta\l-v\delta t$, i.e. it doesn't depend on the potential $V(\l,t)$ (indeed the term involving $V$ exactly cancels when multiplying \eKcontNgd\ and \eEcontNgd ).

Moreover, we have given a very explicit way to compute the correlation functions.
This method (the generalized Darboux Christoffel theorem) might allow one to compute the correlation functions even for $N$ finite, provided one knows the orthogonal polynomials.
What we have computed here is an approximation in the large $N$ limit, using the WKB approximations of the orthogonal polynomials.
This method has shown that the function $x(\l)$ plays a very important role, as the best auxilary variable, its physical meaning should be better understood, beyond the orthogonal polynomial's method.

For instance, another way of studying the chain of matrices is through the loop equations (the invariance of the partition function w.r.t the change of variables, see Appendix D), which would allow to compute the following subleading terms in the $1/N$ expansion.
The loop equations in the large $N$ limit give some algebraic equation (of very high degree) for the resolvent of the first (or last) matrix of the chain.
The parametrization of $\l$ and $\om(\l)$ as functions of $x$ allow to solve this algebraic equation.

\vskip 1cm
\noindent {\bf Acknowledgements}

This work was partly supported by the TMR Network contract ERBFMRXCT 960012.

\vfill\eject


\appendix{A : Correlation functions of any set of eigenvalues}

The general correlation function of any set of eigenvalues has been computed by \rMehtaEyn .

In addition to the kernels $K_{ij}$ and $E_{ij}$, we will need to introduce the $H_{ij}$'s :
$$  H_{ij} = K_{ij} - {1\over N} E_{ji}  \qquad {\rm if }\,\, j\leq i \qquad , \qquad  H_{ij} = K_{ij} \qquad {\rm if }\,\, j>i $$

To compute a partial density of eigenvalues, for instance
$$\rho_{n_0,\dots,n_p}(\l_{0;1},\dots,\l_{0;n_0},\ldots,\l_{p;1},\dots,\l_{p;n_p})$$
 you have to consider all the possible permutations of all the $n=n_0+\dots+n_p$ variables, and then (with obvious notations):
$$ \rho(\l_1,\dots,\l_n)=\mathop{\det}_{i,j} H_{i,j}=\sum_\sigma (-1)^\sigma \prod_i \, H_{i,\sigma(i)}(\l_i, \l_{\sigma(i)}) $$
The connected correlations are obtained by reducing the sum to the cyclic permutations only.

For instance, we have (we assume $i\leq j\leq k$):
$$ \rho_i(\l_i)=K_{i,i}(\l_i,\l_i) $$
$$ \rho^{(c)}_{i,j}(\l_i,\l_j)= -K_{i,j}(\l_i,\l_j)K_{j,i}(\l_j,\l_i) + {1\over N}K_{i,j}(\l_i,\l_j)E_{i,j}(\l_i,\l_j) $$
$$ \rho^{(c)}_{i,j,k} = K_{ij}K_{jk}K_{ki}+K_{ik}K_{kj}K_{ji} - {1\over N}( K_{ij}K_{jk}E_{ik} + K_{ji}K_{ik}E_{jk} + K_{ik}K_{kj}E_{ij} ) +{1\over N^2} K_{ik}E_{ij}E_{jk} $$


\appendix{B : Explicit expressions of orthogonal polynomials}

\noindent The orthogonal polynomial $\P_n$ and $\t{\P}_n$ have explicit expressions as matrix integrals:

$$ \P_n(\l)= \int \d{M_0}_{n\times n}\dots\d{M_{p}}_{n\times n} \, \det{(\l-M_0)}\, \prod_{i=0}^{p}\, \ee{-N\tr V_i(M_i)} \,\,\prod_{i=0}^{p-1}\,\ee{-Nc\,\tr (M_i -M_{i+1})^2} 
 $$

$$ \t{\P}_n(\t\l)= \int \d{M_0}_{n\times n}\dots\d{M_{p}}_{n\times n} \, \det{(\t\l-M_p)}\, \prod_{i=0}^{p}\, \ee{-N\tr V_i(M_i)} \,\,\prod_{i=0}^{p-1}\,\ee{-Nc\,\tr (M_i -M_{i+1})^2} 
 $$

\vfill\eject
\appendix{C: An explicit example: the gaussian potential,  $V(\l)={1\over 2}g\l^2 $ }

\subsec{Discrete case}

Let us consider a constant gaussian potential: $ V'_i(\l)=g\l $. We will use the following parametrization $g=4c (\cosh\chi -1 )$ for later convenience.

The operators $\hat{\l}_i$ contain only two terms, which we write as:
$$ \hat{\l}_i=\alpha_i \hat{x} \sqrt{\hat{n}\over N} + \beta_i \sqrt{\hat{n}\over N} {1\over \hat{x}} $$
The equation of motions are linear and thus easily solved:
$$ \alpha_{i+1}+\alpha_{i-1}-2\alpha_i= {g\over 2c}\alpha_i \qquad , \qquad \beta_{i+1}+\beta_{i-1}-2\beta_i= {g\over 2c}\beta_i $$
with the boundary conditions:
$$ \alpha_0=\beta_p=\alpha \quad ,\quad \alpha_1=(1+{g\over 2c})\alpha \qquad ,\qquad \beta_{p-1}=(1+{g\over 2c})\alpha $$
$$ (1+{g\over 2c})\alpha_p - \alpha_{p-1} = {1\over 2c\alpha} $$
The first line implies that $\beta_i=\alpha_{p-i}$.
The solution is then
$$ \alpha_i=\alpha {\cosh{(i+{1\over2})\chi} \over \cosh{{1\over2}\chi}} \qquad {\rm where}\qquad \alpha^2={1\over 4c\, \sinh{(p+1)\chi} \,\tanh{\chi\over2}} $$
\medskip
The momentum $\hat{P}_i={1\over N}{\d{}\over\d{\l_i}}$ is:
$$ \hat{P}_i^\dagger = -2{1-A_i\over a_i^2} \hat{\l}_i^\dagger + {1\over \alpha_i} \sqrt{\hat{n}\over N} {1\over \hat{x}}
\qquad , \qquad
\hat{P}_i = -2{1+A_i\over a_i^2} \hat{\l}_i + {1\over \alpha_{p-i}} \hat{x}\sqrt{\hat{n}\over N} $$
where we have introduced $A_i={\sinh{(p-2i)\chi}/\sinh{(p+1)\chi}}$ and
$$ a_i = 2\sqrt{\alpha_i\alpha_{p-i}} $$
which will play a very important role as the natural scale for $\l_i$.
Actually $2a_i$ is the width of the distribution of eigenvalues of the $i^{\rm th}$ matrix.
Indeed, in the large $N$ limit and $n\sim N$, we can eliminate $x$ and write $P_i$ as a function of $\l_i$:
$$ P_i(\l_i) = -{2\over a_i}\left( A_i {\l_i\over a_i}  -  \sqrt{{\l_i^2\over a_i^2} -1}\,\right) $$
its imaginary part is the density of eigenvalues of the $i^{\rm th}$ matrix:
$$ \rho_i(\l_i) = {2\over \pi a_i} \sqrt{1- {\l_i^2\over a_i^2}} $$
which is a semi-circle law of diametre $2a_i$.

\medskip
The wave functions can be exactly computed in terms of Hermite polynomials $H_n(x)$:
$$ \psi_{i;n}(\l_i) = c_i\, \left({\alpha_{p-i}\over \alpha_i}\right)^{n\over 2} {1\over\sqrt{n!}} H_n\left( 2\sqrt{N}{\l_i\over a_i}\right) \ee{-N(1-A_i) {\l_i^2\over a_i^2}} $$
$$ \t\psi_{j;n}(\l_j) = \t{c}_j \left({\alpha_{j}\over \alpha_{p-j}}\right)^{n\over 2} {1\over\sqrt{n!}} H_n\left( 2\sqrt{N}{\l_j\over a_j}\right) \ee{-N(1+A_j) {\l_j^2\over a_j^2}} $$
(where $c_i$ and $\t{c}_j$ are some constants, irrelevant for what we need, we just know that $c_i \t{c}_i = {1\over a_i}\sqrt{2N\over \pi}$).

And, we can write some WKB approximations in the large $N$ limit:
$$ \psi_{i;n}(\l_i) \sim \left({\pi\over Nc}\right)^{{i\over2}-{p\over4}}  {2(2\pi)^{-1\over 4} \over\sqrt{a_i \sin{\phi_i}}} \left({\alpha_{p-i}\over \alpha_i}\right)^{2n-2N+1\over 4} \cos{\left[(n+{1\over2})\phi_i - {N\over2}\sin{2\phi_i} -{\pi\over4} \right]} \ee{N A_i {\l_i^2\over a_i^2}} $$
$$ \t\psi_{j;n}(\l_i) \sim \left({\pi\over Nc}\right)^{{p\over4}-{j\over2}}  {2 (2\pi)^{-1\over 4} \over\sqrt{a_{j} \sin{\phi_j}}} \left({\alpha_{j}\over \alpha_{p-j}}\right)^{2n-2N+1\over 4} \cos{\left[(n+{1\over2})\phi_j - {N\over2}\sin{2\phi_j} -{\pi\over4} \right]} \ee{- N A_{j} {\l_j^2\over a_j^2}} $$
where $\l_i=a_i\cos{\phi_i} $.

\medskip
Now let us compute the kernels:

\noindent The propagator $E_{ij}$ is:
$$ E_{i,j}(\l_i,\l_j)= \left({\pi\over Nc}\right)^{j-i-1\over2} \sqrt{\sinh\chi \over \sinh{(j-i)\chi} } \,\ee{-Nc{\sinh\chi \over \sinh{(j-i)\chi}} \left[ \cosh{(j-i)\chi}\, (\l_i^2+\l_j^2) \, - 2 \l_i\l_j \right] } $$
This is an exact result since the saddle point method is exact in the gaussian case.
We also have the determinant of the second derivatives of the potential $D_{ij}(\l_i,\l_j)$ which is a constant:
$$ D_{ij}= {\sinh{(j-i)\chi} \over \sinh\chi } $$

The Kernel $K_{ij}(\l_i,\l_j)$ obeys a generalization of the Darboux Christoffel Theorem:
$$ \left\{ 2{\l_i\over a_i} - 2\tau {\l_j\over a_j}  +  (\tau-{1\over \tau})\left( (1+A_j){\l_j\over a_j} +{a_j\over 2N}{\partial\over\partial \l_j} \right) \right\} K_{ij} = \sqrt{\alpha_i\over\alpha_{p-i}}\psi_{i;N}\t\psi_{j;N-1} - \sqrt{\alpha_{p-i}\over\alpha_i} \psi_{i;N-1}\t\psi_{j;N}$$
$$ \left\{ 2\tau{\l_i\over a_i} - 2{\l_j\over a_j}  -  (\tau-{1\over \tau})\left( (1-A_i){\l_i\over a_i} +{a_i\over 2N}{\partial\over\partial \l_i} \right) \right\} K_{ij} = \sqrt{\alpha_j\over\alpha_{p-j}}\psi_{i;N}\t\psi_{j;N-1} - \sqrt{\alpha_{p-j}\over\alpha_j} \psi_{i;N-1}\t\psi_{j;N}$$
where $\tau$ denotes the ratio $\tau=\sqrt{\alpha_{p-i}\alpha_j / \alpha_i\alpha_{p-j} }$.
Let us emphasize that these equations are exact, even for $N$ finite (remark that when $i=j$ we have $\tau=1$ and we recover the usual Darboux-Christofel theorem).
With the operatorial notations, they are obvious, they are just the rewritting of
$$ \left\{ \l_i(\hat{x}) - \l_i(\hat{y}) \right\} K_{ij} = \left\{ \l_i(\hat{x}) - \l_i(\hat{y}) \right\}{\hat{y}\over \hat{x}-\hat{y}} \psi_{i;N}\t\psi_{j;N}$$
$$ \left\{ \l_j(\hat{x}) - \l_j(\hat{y}) \right\} K_{ij} = \left\{ \l_j(\hat{x}) - \l_j(\hat{y}) \right\}{\hat{y}\over \hat{x}-\hat{y}} \psi_{i;N}\t\psi_{j;N}$$
(if you want to rederive them, be carrefull that here, the partial derivative with respect to $\l_i$ carries on the bra, it is a $P_i^\dagger$).

In the large $N$ limit, we have
$$\eqalign{
K_{ij}\sim &
\left({\pi\over Nc}\right)^{i-j\over2}\, {1\over \sqrt{ a_i\sin{\phi_i}\, a_j\sin{\phi_j} }}\,\,  {\sqrt{\tau}\over 4\pi N}
\,\,\ee{N(A_i\cos^2{\phi_i} - A_j \cos^2{\phi_j})}   \cr
& \qquad \qquad { (1+{1\over\tau}) \sin{ {1\over2}(\phi_i-\phi_j)} \sin{N(\zeta_i-\zeta_j)}
\,\,+\,\, (1-{1\over\tau}) \cos{{1\over2}(\phi_i-\phi_j)} \cos{N(\zeta_i-\zeta_j)}
\over {1+\tau^2\over 2\tau} - \cos{(\phi_i-\phi_j)}}    \cr
& + \quad ({\rm same\,\, thing\,\, with \,\,} \phi_j \to -\phi_j )
} $$
where we have noted $\zeta(\phi)=\phi -{1\over2}\sin{2\phi}$.

\medskip
\noindent - For $i=j$ we have $\tau=1$ and we recover:
$$ K_{ii}(\l,\mu) \sim {1\over a_i \sqrt{ \sin{\phi}\, \sin{\t\phi} }}\,  { 1\over 4\pi N}\, \left( {\sin{N(\zeta(\l)-\zeta(\mu))}\over \sin{\phi-\t\phi\over2}} \right) \,\ee{{N\over 2} {A_i\over a_i^2} ( \l^2 - \mu^2 ) }$$
which gives for $\l=\mu$ the density of eigenvalues:
$ \rho_i(\l)= {1\over \pi a_i}\sin\phi $.

\medskip
\noindent - When $i\neq j$, we remark that the denominator never vanishes.
It is maximum when $\phi_i=\phi_j$, i.e. $\l_i/a_i = \l_j/a_j$, this is the regime in which one could have a not too small correlation.


\subsec{Continuous limit}

It can be obtained by setting $c={1\over 2\epsilon}$, $t=i\epsilon$, $T=p\epsilon$, $g=\epsilon \nu^2$, and taking the limit $\epsilon\to 0$.

We find:
$$ \alpha^2= {1\over \nu\sinh{\nu T}} $$
$$ \l(x,t)=\alpha(x\cosh{\nu t} + {1\over x} \cosh{\nu (T-t)} ) $$
which we write
$$ \l(\phi,t) = a(t) \cos\phi $$
$a(t)$ is given by
$$ a(t)=2\alpha \sqrt{ \cosh{\nu t} \cosh{\nu (T-t)} } $$

\medskip
In the small distance regime ($\delta \l$ small and $\delta t$ small), we have (up to some factors which will cancel when we compute the correlation function):
$$ K(\l,t | \l+\delta\l , t+\delta t) \sim {1\over 2\pi N}{1\over a\sin\phi} {\delta\phi \sin{(N\pi\rho a\sin\phi \delta\phi)} + {2\over a^2} \delta t \cos{(N\pi\rho a\sin\phi \delta\phi)} \over 2{\delta t^2\over a^4} + {1\over 2}{\delta \phi^2}}$$
$$ {1\over N} E(\l,t | \l+\delta\l , t+\delta t) \sim {1\over \sqrt{2\pi N \delta t}} \ee{-{N\over 2}(a^2\sin^2\phi {\delta \phi^2\over \delta t} + {4\over a^2} \cos^2\phi \delta t)}$$
And thus we have the two-point connected correlation function:
$$ \rho^{(c)}(\l,t | \l+\delta\l , t+\delta t) = - K(\l,t | \l+\delta\l , t+\delta t) \left( K(\l+\delta\l , t+\delta t | \l, t) - {1\over N} E(\l,t | \l+\delta\l , t+\delta t) \right) $$

which gives:
\eqna\ecorelgaus
$$ \eqalign{
 \rho^{(c)} = & - {1\over 4\pi^2 N^2 a^2 \sin^2\phi} { \delta\phi^2 \sin^2{(N\pi\rho a\sin\phi \delta\phi)} - {4\over a^4} \delta t^2 \cos^2{(N\pi\rho a\sin\phi \delta\phi)} \over (2{\delta t^2\over a^4} + {1\over 2}{\delta \phi^2})^2}  \cr
& + {1\over ( 2\pi N \delta t)^{3/2}} {\delta t \over a\sin\phi} {\delta\phi \sin{(N\pi\rho a\sin\phi \delta\phi)} + {2\over a^2} \delta t \cos{(N\pi\rho a\sin\phi \delta\phi)} \over 2{\delta t^2\over a^4} + {1\over 2}{\delta \phi^2}} \cr
& \hskip 4cm\ee{-{N\over 2}(a^2\sin^2\phi {\delta \phi^2\over \delta t} + {4\over a^2} \cos^2\phi \delta t)} \cr
}\ecorelgaus{}$$

\appendix{D: The loop equations}

The loop equations are just  the consequence of the invariance of the partition function
$$ Z= \int \d{M_0}\dots\d{M_p} \, \ee{-N\tr \sum_i V_i(M_i) + c\sum_{i} (M_i-M_{i+1})^2 } $$
with respect to some change of variables
$ M_i \lra M_i + \epsilon f(M_0,\dots,M_p) $.
They are most simply written by introducing the resolvant
$$\om(z)={1\over N}\left< \tr {1\over z-M_0} \right> $$
Actually, one can easily obtain close equations in the large $N$ limit only for the resolvant of the first ($i=0$) and last ($i=p$) matrices.
We will give here the recipe to write the algebraic equation for $\om(z)$ but without any proof (the proof obtained by multiple recurrences is quite tedious but not complicated):

first we define the functions:
$$ z_0(z) = z \qquad , \qquad z_1(z) = z + {1\over 2c}V'_0(z) - {1\over 2c}\om(z) $$
$$ {\rm and} \qquad z_{i+1}(z) = 2z_i(z) + {1\over 2c}V'_i(z_i(z)) - z_{i-1}(z) \qquad {\rm for} \quad 0<i<p $$
each function $z_i(z)$ is a polynomial in $z$ and $\om(z)$.
The algebraic equation relating $z$ to $\om(z)$ is then:
\eqn\eloop{ \left( z_0(z)-z_{1}(z)+{1\over 2c}V'_0(z_0(z)) \right)\left( z_p(z)-z_{p-1}(z)+{1\over 2c}V'_p(z_p(z)) \right) = {1\over 2c} P(z_0(z),\dots,z_p(z)) }
where the right hand side $P(x_0,\dots,x_p)$ is a polynomial of $p+1$ variables (now considered as independent):
$$ P(x_0,\dots,x_p) = {\rm Pol}_{x_0,\dots,x_p} \,  {1\over N} \left< \tr\, f_{p+1}(x_0,\dots,x_p) {1\over x_0-M_0}\dots{1\over x_p-M_p}\right>$$
and the functions $f_n(x_0,x_2,\dots,x_{n-1})$ are defined as follows:
$$ f_0 =1 \qquad , \qquad f_1(x)= x+{1\over 2c}V'_0(x) $$
$$ f_{n+1}(x_0,\dots,x_{n})=f_n(x_0,\dots,x_{n-1})(x_{n}+{1\over 2c}V'_n(x_{n})) - f_{n-1}(x_0,\dots,x_{n-2}) x_n x_{n-1} $$
Note that all this is symetric in the exchange of the two extremities of the chain $i\to p-i$.

Most of the coefficients of the polynomial $P$ are unknown, they are some arbitrary constants which have to be fixed by some analytical considerations about $\om(z)$.

\noindent - One of the constraints is that $\om(z)\sim {1/z}$ when $z\to\infty$.

\noindent - The other constraints require more physical input.
For a single well potential $V_i(\l)$, we expect the density of eigenvalue to have a connected support, i.e. that $\om(z)$ has only one-cut.

This one-cut asumption should allow in principle to determine all the unknown coefficients of $P$.
We made exactly the same asumption when we replaced the operator $\hat{x}$ by a number $x$ in the large $N$ limit, and we observe that the functions $\l_i(x)$ give a parametrization of the solution of eq \eloop .
Actually, if $z=\l_0(x)$ we have that each $z_i(z)$ is nothing but $\l_i(x)$, and the loop equation \eloop\ is nothing but the product of \einiO\ and \einiM .

\listrefs

\bye